\documentclass[aps, twocolumn, showpacs,notitlepage]{revtex4-1}
\usepackage{amssymb}
\usepackage{graphics}  

\usepackage{amsfonts}
\usepackage{amsmath}
\usepackage{amssymb}
\usepackage{graphicx}
\usepackage{color}

\newcommand{\marge}[1]{\marginpar{}}  
\newcommand{\Sl}[1]{{}}           
\newcommand{\beq}[1]{\Sl{#1}\begin{equation}\if#1\empty\else\label{#1}\fi}
\newcommand{\eeq}{\end{equation}}
\newcommand{\beqa}{\begin{eqnarray}}
\newcommand{\eeqa}{\end{eqnarray}}
\newcommand{\beal}{\begin{align}}
\newcommand{\enal}{\end{align}}

\definecolor{red}{rgb}{1,0,0}

\providecommand{\U}[1]{\protect\rule{.1in}{.1in}}
\begin{document}

\title{Comment on "Possible divergences in Tsallis' thermostatistics"}
\author{Jean Pierre Boon and James F. Lutsko}
\affiliation{Physics Department, CP 231, Universit\'e Libre de Bruxelles, 1050 - Bruxelles, Belgium}
\email{jpboon@ulb.ac.be}
\homepage{ http://homepages.ulb.ac.be/~jpboon/}
\email{jlutsko@ulb.ac.be}
\homepage{http://www.lutsko.com}

\pacs{05.20.-y, 05.70.Ce.Gg, 05.90.+m} 


\begin{abstract}
The problematic divergence of the $q$-partition function of the harmonic oscillator
recently considered in \cite{plastino} is a particular case of the non-normalizabilty of
the distribution function of classical Hamiltonian systems in non-extensive thermostatistics
as discussed previously in \cite{lutsko-boon}.
\end{abstract}

\date{\today}
\maketitle

\bigskip


In a recent paper \cite{plastino} Plastino and Rocca discuss the problematic divergence
occurring in the computation of the partition function of the harmonic oscillator 
in the non-extensive thermostatistics formulation. The problem is that the 
normalization of the distribution function leads to a diverging quantity as briefly
summarized below.

The general formulation of non-extensive statistical mechanics is developed
on the basis of three axioms: (i)~the $q$-entropy for systems with
continuous variables is given by~\cite{tsallis88} 
\begin{equation}
S_{q}\,=\,k_{B}\,\frac{1-{K\int \rho ^{q}\left( \Gamma \right) d\Gamma }}{q-1%
}\,,  \label{STq}
\end{equation}%
where $\Gamma$ is the phase space variable and
$K$ must be a quantity with the dimensions of $\left[ \Gamma \right]
^{q-1}$, i.e. $K=\hbar ^{DN\left( q-1\right) }$
($D$ denotes the dimension of the system and $N$ the number of particles)
and the classical Boltzmann-Gibbs entropy is retrieved in the limit $q\rightarrow 1$; 
(ii)~the distribution function $\rho \left( \Gamma \right) $ is slaved to the
normalization condition 
\begin{equation}
1\,=\,\int \rho \left( \Gamma \right) d\Gamma \,;  \label{Norm}
\end{equation}%
(iii)~the internal energy is measured as 
\begin{equation}
U\,=\,{\int P_{q}\left( \Gamma \right) H\,d\Gamma }\,=\,\frac{\int \rho %
^{q}\left( \Gamma \right) H\,d\Gamma }{\int \rho ^{q}\left( \Gamma \right) %
d\Gamma }\,,  \label{Uesc}
\end{equation}%
where $P_{q}\left( \Gamma \right) $ is the escort probability distribution~%
\cite{beck_schlogl} which is the actual probability measure.   

The distribution function $\rho \left( \Gamma \right) $ is obtained by
maximizing the $q$-entropy subject to the normalization (\ref{Norm}) and
average energy (\ref{Uesc}). For the harmonic oscillator with Hamiltonian
$H = P^2 + Q^2$, the classical partition function (using the notation in \cite{plastino})
\beqa
{\cal Z} &=& \,\int_{-\infty}^{\infty} \, \rho ( {\tt p, q}) \,d^{N} {\tt p} \, d^{N} {\tt q} \nonumber\\
&=& \,\int_{-\infty}^{\infty} {\exp{\left(-\beta \left( P^{2} + Q^{2} \right)\right)}}\,%
d^{N} {\tt p} \, d^{N} {\tt q} 
\label{NormClas}
\eeqa%
(in $N$-dimensional $({\tt p,q})$ space) is replaced by its non-extensive analogue, 
i.e. the exponential function is replaced by a $q$-exponential \cite{tsallis09}, giving
\begin{equation}
{\cal Z}\,=\,\frac{\pi^{N}}{\Gamma (N)}\,\int_{-\infty}^{\infty} U^{N-1}%
\left( 1-\left( 1-q\right) \beta U \right) ^{\frac{1}{1-q}}\,dU \,,
\label{NormNext}
\end{equation}%
where $U = P^{2} + Q^{2} $ is the harmonic oscillator energy. The authors note that 
"the partition-defining integral diverges for $q\geq 3/2$ and $N \geq 3$ ".

The non-normalizibility problem has been examined in detail in \cite{boon-lutsko} 
for the $q$-ideal gas and more generally in \cite{lutsko-boon} thereby questioning 
the validity of non-extensive thermodynamics for Hamiltonian systems. 
The analytical results in  \cite{lutsko-boon}  show that
(i)~the non-extensive thermodynamics formalism should be called into
question to explain experimental results described by extended 
distributions exhibiting long tails, i.e. $q$-exponentials with $q>1$, and that
(ii)~in the thermodynamic limit the theory is only consistent in the
range $0\leq q\leq1$ where the distribution has finite support, thus
implying that configurations with e.g. energy above some limit have zero
probability. In particular it was shown in \cite{lutsko-boon} that for $q>1$, 
the resulting distributions only exist for the restricted range 
$0<q<1+{\mathcal O}\left( \frac{1}{N}\right) $.
Since the so-called "fat-tailed" distributions correspond to $q>1$, this
means that generalized $q$-thermodynamics cannot be seen as an explanation of
the occurrence of such distributions in Hamiltonian systems with $N$ large
\cite{non-ext}.

The harmonic oscillator  belongs to the class of systems which are considered
in \cite{boon-lutsko} and  in \cite{lutsko-boon}. The divergence obtained 
in \cite{plastino} is essentially a consequence of the quadratic form of the
Hamiltonian and therefore appears as a particular case of the systems
discussed in \cite{lutsko-boon} where it was shown that in fact, independent of the form of the potential,  the quadratic kinetic energy term is enough to impose the limits quoted above.

The authors in \cite{plastino} offer a possible remedy by using the $q$-Laplace
transform. One can indeed define $q$-functions as generalized analogues of
classical functions whose mathematical object involves an exponential form such as
e.g. the Fourier and Laplace transforms. However replacements (such as 
$\exp \rightarrow \exp_q$) imply a complete mathematical reformulation of statistical 
mechanics beyond the program proposed by $q$-thermostatistics \cite{tsallis09}
and whose physical basis remains to be justified. In particular, it is not clear how one might write the partition function in the form of a Laplace transform for any system other than the harmonic oscillator or ideal gas. It is therefore not apparent how this program could be implemented in general.

\begin{acknowledgments}
The work of JFL was supported by the European Space Agency under contract number
ESA AO-2004-070.
\end{acknowledgments}


\end{document}